\documentclass{elsart}
\usepackage{graphics}
\usepackage{epsfig}
\usepackage{amssymb}
\begin{document}

\begin{frontmatter}

\title{Integer Quantum Hall Effect in Graphite}

\author{H. Kempa\thanksref{kem}},
\thanks[kem]{Present address: Institute for Print and Media Technology
at Chemnitz University of Technology, Reichenhainer Str. 70, D-09126
Chemnitz, Germany}
\author{P. Esquinazi}\ead{esquin@physik.uni-leipzig.de}, and
\author{Y. Kopelevich\thanksref{yako}}
\thanks[yako]{On leave from Instituto de Fisica, Unicamp,
13083-970 Campinas, Sao Paulo, Brasil.}
\address{Division of Superconductivity
and Magnetism,  Institut  f\"ur Experimentelle Physik II,
Universit\"at Leipzig, Linn\'estr. 5, D-04103  Leipzig, Germany}

\begin{abstract}
We present Hall effect measurements on highly oriented pyrolytic
graphite that indicate the occurrence of the integer
quantum-Hall-effect.  The evidence is given by the observation of
regular plateau-like structures in the field dependence  of the
transverse conductivity obtained in van der Pauw configuration.
Measurements with the Corbino-disk configuration support this
result and indicate that the quasi-linear and non-saturating
longitudinal magnetoresistance in graphite is governed by the Hall
effect in agreement with a recent theoretical model for disordered
semiconductors.
\end{abstract}
\end{frontmatter}

 Recent experimental\cite{1,yakovadv03,ocana03,ulrich04,luky04,matsui05}
and theoretical\cite{2,3,4} work demonstrates the occurrence of correlated
phenomena in the electron system of graphite, indicating that the
classical view of the physics of its magnetotransport properties is not
adequate. Besides, the quantum Hall effect (QHE) \cite{1} as well as
evidence for Dirac fermions (massless particles due to the linear
dispersion relation) \cite{luky04} have been reported for  bulk samples of
highly oriented pyrolytic graphite of high quality. Recently published
 results obtained on bulk graphite \cite{041,051}, a few layers thick graphite \cite{science}, and
  in graphene (single graphite layer) \cite{novo05,zhang05}
have confirmed both observations and showed that in graphene the QHE is
anomalous in that the contribution of the Dirac fermions makes the
plateaus to occur at half-integer filling factors. Certainly, the
understanding of the electronic properties of graphite is necessary for
the development of advanced technology of the graphite-related
nanomaterials.

The experimental work we report here is based partially on data presented
in an international meeting \cite{carbon04}. These data clearly
demonstrate the quantization of the Hall effect and in particular  the
existence of an even QHE in graphite, i.e. a transverse voltage that is
symmetric on magnetic field reversal. The occurrence of the QHE in bulk
graphite indicates that the coupling between graphene layers is much
smaller than the one assumed by the Slonczewski-Weiss-McClure theory
\cite{swm} but perfectly agrees with the results obtained by Haering and
Wallace \cite{wal}. Also, the experimental results we present here
indicate a possible solution to the longstanding problem of the
quasi-linear and non-saturating magnetoresistance (MR)\cite{5} providing
evidence that the MR at high enough fields is governed by the Hall
resistance, in agreement with a recent theoretical model \cite{6}.

The occurrence of the QHE is not restricted to perfectly two-dimensional
(2D) systems, as measurements in highly anisotropic layered systems
\cite{7}, organic Bechgaard salts \cite{8}, Molybdenum bronzes \cite{9},
layered crystals of the type (Bi$_{0.25}$Sb$_{0.75})_2$T$_3$ \cite{10},
(TMTSF)$_2$AsF$_6$ system \cite{11}, as well as in 200-layer quantum-well
structures \cite{12} indicate. Nevertheless, it is surprising that for the
paradigm of strongly anisotropic systems, graphite, no clear evidence for
the QHE has been published till 2003 \cite{1,yakovadv03,ocana03}.  The
main obstacle to obtain clear evidence for the QHE in graphite is the
sample quality, i.e. the internal short circuits between graphene planes
caused by defects and impurities. As a characterization of the sample
quality and orientation of the samples one may use the full width at half
maximum (FWHM) of the rocking curves. In general we observed that the
smaller the FWHM and the distance between voltage electrodes, the clearer
are the plateaus in the Hall effect. The samples measured in this work are
highly oriented pyrolytic graphite (HOPG) obtained from Advanced Ceramics
(AC sample) and from the research institute "Graphite" in Moscow (sample
HOPG-3), with FWHM equal to 0.40$^\circ$ and 0.64$^\circ$, respectively.

Angle dependent magnetoresistance (MR) oscillations as well as a ratio of
the order of $10^4$ between $c$-axis and in-plane resistivities in highly
oriented graphite samples\cite{yakovadv03} indicate a much larger
anisotropy, a much smaller coupling between the graphene layers and a
greater influence of the two-dimensionality of the sample on the transport
properties than earlier studies have reported. Among these findings is the
magnetic-field induced metal-insulator-transition  (MIT), which shows
significant similarities to the one discovered in dilute 2D electron
systems \cite{14}. As in graphite \cite{1},  in some of these 2D systems
the MITs have been shown to be connected to
quantum-Hall-insulator-transitions at high magnetic fields \cite{15}.
After these experimental observations and the first theoretical prediction
for the QHE in graphite \cite{16}, which was followed by further developed
theoretical studies \cite{gus05,per05}, the occurrence of
quantum-Hall-states in graphite was already expected.

The longitudinal resistance as a function of magnetic field of all samples
has been measured with the conventional ac-method, where four electrodes
made of silver paint are placed on top of a surface parallel to the
graphene layers. To calculate from this measurement the absolute
resistivity (longitudinal as well as transverse) it is necessary to know
the effective penetration depth of the current $\lambda$. Independent
measurements at room temperature using electrodes at different positions
of the same sample reveal that for macroscopic, bulk samples $\lambda$ is
in the range $1~\mu$m$ \lesssim \lambda \lesssim  10~\mu$m. With a
conventional electrode geometry a good estimate of the resistivity is
given by $\rho \sim ~ b \lambda V/Id$ where $b$ is the width of the
sample, $I$ the applied electrical current,  and $V$ and $d$ are the
voltage and distance between the inner contacts. However, part of the data
presented here were obtained using a van der Pauw configuration consisting
of four point-like electrodes ($1\ldots 4$) placed at the corners of the
main sample surface. In the van der Pauw configuration the typical surface
between contacts fixed at the corners of a parallelepiped was $\sim
10~$mm$^2$. If we apply a current $I_{12}$ between contacts 1 and 2 and we
measured the voltage $V_{34}$ between contacts 3 and 4, the longitudinal
resistivity is given by the equation $\rho_{xx} = 4.53 \lambda
V_{34}/I_{12}$ \cite{pauw}. The transverse resistivity is obtained from
this arrangement measuring the voltages $V_{42}$ and $V_{13}$ with the
input currents $I_{13}$ and $I_{42}$, respectively, using $\rho_{xy} = 0.5
\lambda (V_{42}/I_{13} - V_{13}/I_{42})$ \cite{pauw}. The longitudinal and
transverse conductivities $\sigma_{xx}$ and $\sigma_{xy}$ are calculated
from the measured resistivities inverting the two dimensional matrix, i.e.
\begin{equation}\label{sig}
    \sigma_{xx} = \frac{\rho_{xx}}{\rho_{xx}^2 + \rho_{xy}^2}\,\,\,\,,\,
    \sigma_{xy} = \frac{\rho_{xy}}{\rho_{xx}^2 + \rho_{xy}^2}\,.
\end{equation}
The magnetic field was always applied perpendicular to the graphene layers
(the main sample surfaces). For all geometries the applied current was of
the order of 1~mA,  which is in the ohmic regime and for which no heating
effects are observed. The resistance was measured with a LR-700 resistance
bridge from Linear Research Inc., which works at low ac-frequency.

The essential signature of the QHE is the occurrence of plateaus in the
Hall conductivity $\sigma_{xy}(H)$ or resistivity $\rho_{xy}(H)$
isotherms. The transverse conductivity value at the position of the
apparent first plateau-like feature at $H \sim 5$T$/\mu_0$ is
 for this sample (HOPG-3) \cite{1,yakovadv03,ocana03} $\sigma_{xy}(H \sim
5$T$/\mu_0, T \le 2~$K$) \simeq 0.27/\lambda ~ \Omega^{-1}$m$^{-1}$, where
$\lambda$ is given in meters. Assuming an exponential decay of the current
with the distance from the surface and taking into account that the
current penetration depth $\lambda >> c/2$, the distance between the
graphene layers $c/2 = 3.35~\AA$, the contribution of a single graphene
layer can be written as

\begin{equation}
\sigma^\square_{xy}=\sigma_{xy}\lambda\left(1-\exp\left(-\frac{c}{2\lambda}\right)
\right) \simeq \sigma_{xy} c /2\,. \label{sigmasq}
\end{equation}

If we assume that the first plateau occurs when $\sigma^\square_{xy} =
\sigma_{xy} (c/2) = e^2/h = 3.86 \times 10^{-5}~\Omega^{-1}$, using our
data we estimate a penetration depth $\lambda = 2.35~\mu$m, which is
within the expected range. Figure 1 shows both conductivities for the
HOPG-3 sample obtained from the measured resistivities and calculated
using $\lambda = 2.35~\mu$m. This figure shows that regular plateaus are
clearly observable in the $\sigma^\square_{xy}(H)$ (or in
$\sigma_{xy}(H)$) measured in the van der Pauw configuration and plotted
as a function of the inverse applied field. From the data in Fig.1 we
obtain that the transverse conductivity
 for a single layer at the apparent first plateau agrees within experimental error
($\pm 50\%$ due to the uncertainty in the absolute value of $\lambda$)
with the separation between the plateaus $\Delta \sigma^\square_{xy}
\simeq 3 \times 10^{-5}~ \Omega^{-1}$.

\begin{figure}
\centerline{\psfig{file=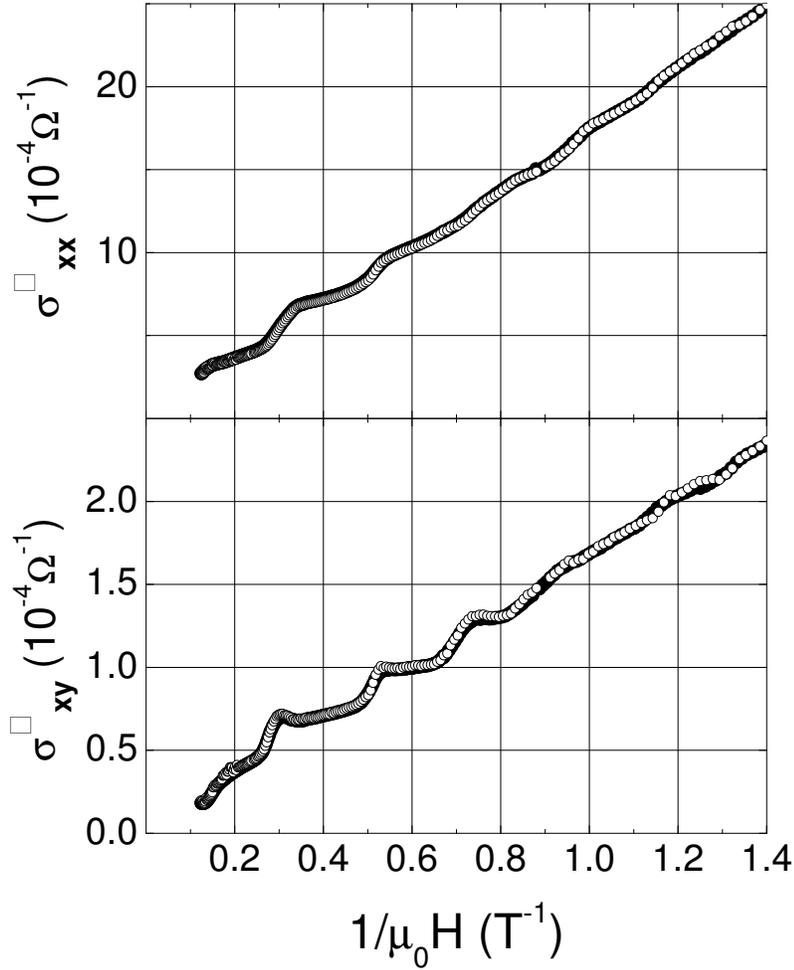,width=11.0cm}} \caption{The components of
the conductivity tensor per graphene layer vs. inverse field for the
sample HOPG-3 measured in van der Pauw-configuration at 0.1 K $(\circ)$
and 2 K $(\bullet)$. The conductivities were calculated using
Eqs.~(\protect\ref{sig}) and (\protect\ref{sigmasq}) and assuming a
penetration depth $\lambda = 2.35~\mu$m. The absolute values of the
conductivities $\sigma_{xx}$ and $\sigma_{xy}$ can be easily obtained
dividing the plotted values by $3.35 \times 10^{-10}$m. Smaller (larger)
values of the penetration depth $\lambda$ increase (decrease) the absolute
values of the conductivities. In the off-diagonal element regular plateaus
are seen, indicative of the quantum-Hall-effect. We note that in this
configuration the transverse Hall voltage changes sign when the field is
reversed.} \label{vdPauw}
\end{figure}

 Considering the simplicity of  the model and the geometrical
 errors involved in the determination of the absolute values
the quantitative agreement with the
 expected value for an integer-like QHE seems reasonable.
 Within experimental error, the value obtained for $\sigma_{xy}$
 depends on sample and also on the electrode arrangement,
 indicating that the disorder and current distribution may
 have an influence. From the sign of the Hall signal we conclude
 that the carriers have the same electronic charge as the electrons.

 A vanishing longitudinal resistivity $\rho_{xx}(H)$ in the vicinity
 of the fields where the Landau levels are filled is a concomitant
 of the QHE, which appears in 2D systems at very low temperatures
 \cite{17}.
 However, $\rho_{xx}(H)$ (or $\sigma_{xx}(H)$) of graphite does not
 show any minima or a decreasing behavior with field, see Fig.1.
 The positive MR of graphite is very large in clear contrast to
 the relatively small MR in most of the 2D systems showing the QHE.
 Moreover, it is nearly linear in field without any sign of saturation (see Fig.1).
 High resolution measurements of the MR of HOPG samples indicate a
 quasi-linear field dependence at fields as low as $\sim 10^{-3}~$T
 with an anomaly in the field exponent at the MIT at $\mu_0 H \sim 0.1~$T
 \cite{18}.  Note that the magnetoconductance at high fields, where the Shubnikov - de Haas (SdH)
 oscillations due to the Landau level quantization occur, shows a striking
  similarity with the transverse conductivity, see Fig. 1.

The origin for the linear MR (LMR) in graphite as well as in other
semimetals is a matter of current discussion \cite{5,6,19}. The simple two
band model \cite{20} does not explain the quasi-LMR observed at low and
high fields, unless one assumes ad-hoc field dependences for the electron
and hole mobility. Abrikosov proposed that the LMR takes place in the
Landau level quantization regime and above the field $H_{QL}$ that pulls
carriers into the lowest Landau level. In order to account for the low
values of the characteristic field ($H_{QL} \sim 10^{-3}~$T), the
graphite-like energy spectrum with a linear dependence of energy on
momentum has been assumed \cite{5,19}. The theory clearly states that LMR
should be a generic property of graphene, which is the model system. The
evidence we provide for an integer-like QHE casts doubts whether the
transport properties of graphite can be appropriately described by this
``quantum magnetoresistance" model added to the fact that a quasi-linear
MR is observed also at low fields.

Experimental evidence based on a large number of transport measurements in
different HOPG samples indicates that the internal disorder of a graphite
sample influences the field and temperature dependence of the transport
properties in such an extent that the QHE can in some cases transform in a
linear field dependent Hall effect with minima at specific fields
\cite{1}. Moreover, different contact distributions on the same sample
show also a small but striking influence on the transport properties.
Therefore these samples should be considered as disordered semimetals.
Regarding the quasi-linear and non-saturating MR in disordered
semiconductors, Parish and Littlewood showed recently that this behavior
might be provided internally by the Hall effect \cite{6}. In what follows
we demonstrate experimentally that the Hall-like signal can be measured in
a graphite sample at large enough fields in a voltage contact
configuration where for the homogeneous case this signal is not expected.

The sample for a Corbino-disk experiment has been prepared by means of
e-beam lithography using a 200~nm gold layer sputtered on the sample (AC)
main surface, as shown in the inset of Fig.~2. In the Corbino disk
geometry the diameter of the outer ring was 1.4~mm and 0.1~mm its width as
well as the diameter of the electrodes. In this configuration, the current
is applied between the inner point-like electrode and four contacts along
the outer ring-like electrode to assure homogenous current input. The
voltage is measured between any of the other contacts. In a homogenous,
isotropic (along the planes) 2D sample, the application of a constant
current $I$ would result in a two-dimensional radial current density $j_r=
I / 2\pi r$ at a point located at distance $r$ from the center, which is
deflected by the magnetic field by the Hall-angle $\theta$ with
$\tan(\theta) = \sigma_{xy} /\sigma_{xx}$. The result is a current path
along logarithmic spirals, consisting of the radial component $j_r$ and an
azimuthal (circulating) component given by \cite{21} $|j_\phi|  =  | j_r
\sigma_{xy} /\sigma_{xx} |$, see the inset in Fig.~2. If we neglect any
influence of the Hall component, the voltage between any two contacts at
the distances $r_1$ and $r_2$ from the center is then given by
\begin{equation}
V=\frac{1}{\sigma_{\square}}\int_{r_1}^{r_2}j_rdr
=\frac{I}{2\pi\sigma_{\square}}\ln\frac{r_2}{r_1}\,. \label{V}
\end{equation}

Here $\sigma_{\square}$    is the longitudinal in-plane conductivity. As
expected, the voltage vanishes between contacts located at the same
distance from the center. The circulating current component is caused by
the Lorentz force perpendicular to the electric field and therefore it is
a non-dissipative current, which does not contribute to the voltage. The
aim of the experiment is to show that when for a given pair of electrodes
inside the Corbino disk the ``longitudinal"-like signal is minimized (i.e.
$r_1 \simeq r_2$, see Eq.~(\ref{V})), still a voltage is measured and this
resembles the Hall signal.

\begin{figure}
\centerline{\psfig{file=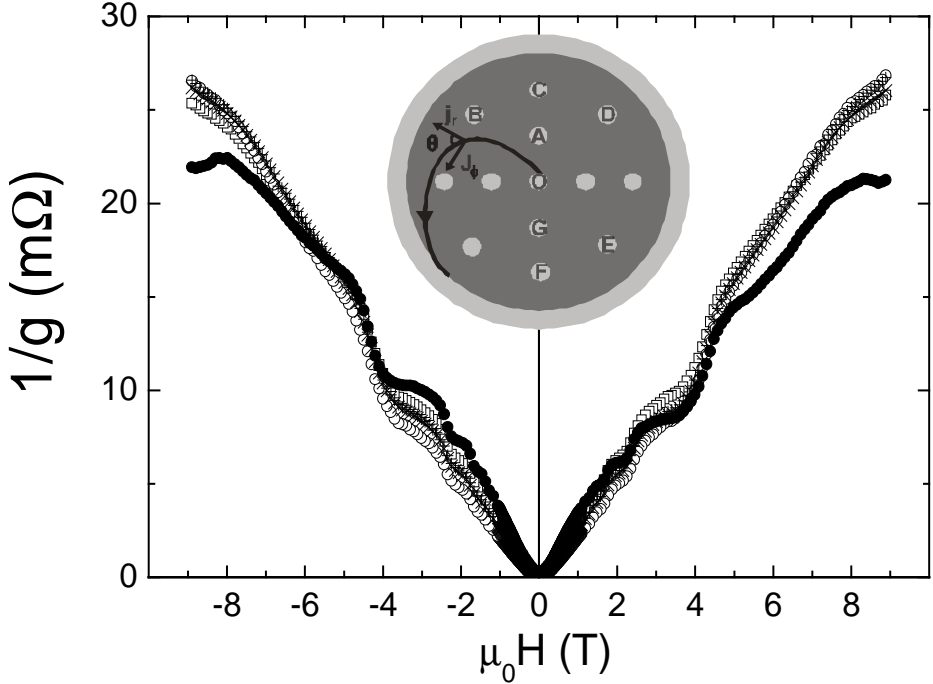,width=15.0cm}} \caption{Inverse
diagonal component of the conductivity tensor vs. applied field for the AC
sample measured in Corbino-configuration at 2 K. The conductance $g \sim
\sigma_{\square}$   is calculated from Eq.(3) assuming a relation of the
distances of the voltage contacts $x=r_2/r_1$ which leads to the best
scaling onto the curve for the contact pair A-C (see below). The current
is applied between the inner electrode and the outer ring; voltage is
measured between pairs of the other electrodes. The curves were obtained
from the voltage contacts: $(+)$ A-C, $x = 1.5$; $(X)$ G-F, $x = 1.957$;
$(\square)$ B-C, $x= 1.111$; $(O)$ E-F, $x= 0.915$; $(\bullet$) C-D,
$x=1.009$. The electrode designations correspond to the ones in the inset.
Inset: View from top onto the Corbino- configuration with designations of
the electrodes as used in the text. The current path along a logarithmic
spiral is schematically shown together with the radial component $j_r$,
the azimuthal component $j_\phi$ and the Hall-angle $\theta$.}
\label{Corbino1}
\end{figure}

Figure 2 shows the field dependence of the inverse conductivity obtained
from voltage measurements between various contact pairs. Two of them (A-C
and G-F) provide the longitudinal magnetoconductivity and three pairs at
nominally the same distance from center (B-C, E-F and C-D) should show no
voltage. In contrast to the expectations for a homogeneous sample, the
voltage between pairs located at nominally the same distance from the
center is not negligible. The first explanation for this observation would
be that this is due to a misplacement of the contacts giving rise to a
radial distance between them. To check this, the data in Fig. 2  is
presented as follows. For the contact pair A-C (see inset in Fig. 2) the
conductance $g \sim \sigma_{\square}$  is calculated from (3) assuming the
nominal value for $x=r_2/r_1=0.45$~mm/0.30~mm=1.5. For the other pairs,
$x$ is chosen in such a way, that the curves scale onto the one for the
pair A-C.

\begin{figure}
\centerline{\psfig{file=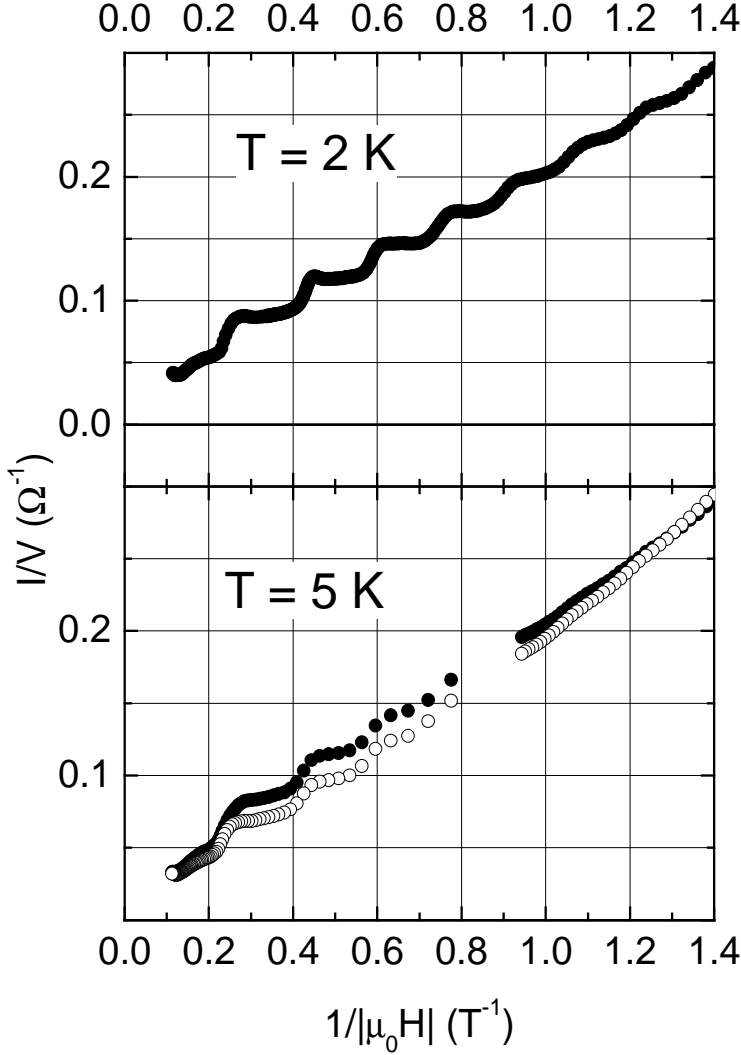,width=12.0cm}} \caption{Data
obtained from the contact pair C-D in the Corbino-configuration.
Top: Curve taken at 2 K. Bottom: Curve taken at 5 K at positive
($\bullet$) and negative ($\circ$) field directions. Note that the
voltage does not change sign when the field is reversed.}
\label{Corbino2}
\end{figure}

As can be seen in Fig. 2 the curve for the pair G-F scales very well
assuming $x=1.957$, corresponding to a shift of the inner contact from the
nominal one by $70~\mu$m, caused by the limited accuracy of the
preparation (nominal electrode diameter is $100~\mu$m). The same holds for
the pairs B-C and E-F, and therefore we conclude that those signals are
mainly determined by the "nominal" longitudinal MR. The pair C-D, however,
has the smallest nominal misplacement ($x=1.009$), but shows the most
significant deviation from the assumable purely longitudinal MR curve A-C.
Interestingly, C-D is the contact pair where we observe the clearest
plateau-like structures similar to the ones measured in the van der Pauw
configuration, as shown in the upper part of Fig. 3. The periodicity of
the plateaus agrees very well with that obtained from the van der Pauw
configuration. The inverse-field period of the plateaus is
$\Delta(1/\mu_0H) \simeq 0.17~$T$^{-1}$. Using the Onsager relation
\cite{22} for a 2D electron system $\Delta(1/\mu_0H) = e/\pi\hbar n_{2D}$
we obtain a 2D electron density for our HOPG samples $n_{2D} \simeq 2.85
\times 10^{15}~$m$^{-2}$. This value is comparable with literature values
for the 3D electron density $n_{3D} \sim 3 \times 10^{24}~$m$^{-3} \sim
n_{2D} 2 / c$. From this evidence we conclude therefore, that the measured
voltage between the electrodes C-D at large enough fields is not related
to the longitudinal MR.

The origin for the unexpected Hall signal at the pair C-D is compatible
with the disordered semiconductor model of Ref.\cite{6}. In this model the
LMR is governed by the Hall resistance of the current paths. Note that due
to the influence of $\rho_{xy}$ in $\rho_{xx}, \sigma_{xy} \sim 1 /
\rho_{xy}$. However, this signal is independent of the field direction
(switching the field direction will also switch the transverse direction
of the current within the disordered network). That means that the
measured MR at C-D should be field direction independent in case the Hall
current paths are perfectly symmetric against field inversion. To check
this we performed measurements at the two field directions, i.e. $+z$ and
$-z$. The results shown in Fig. 2 and in Fig. 3 indicate that the observed
voltage does not change sign and does depend only slightly on the field
direction probably due to non-identical current paths for both field
directions as expected for a real disordered system. The voltage at the
pair C-D at fields below $\sim 0.3~$T indicates that a longitudinal MR
signal mixes into the overall signal showing an anomaly at the MIT at
$\sim 0.1~$T \cite{18}.

We would like to note that an even Hall effect that occurs in
inhomogeneous materials is a well known phenomenon. This effect has been
observed in the mixed state of strongly inhomogeneous  type-II
superconductors \cite{sta,kop,vil}  as well as in inhomogeneous
semiconductors \cite{bul,bar}.

In summary, transport measurements performed in different
configurations on HOPG samples provide evidence for an
integer-like QHE in agreement with theoretical expectations. Our
experiments indicate that the absence of minima in the
longitudinal MR (that usually accompany the plateaus in the Hall
resistivity) as well as the non-saturating LMR observed in all
graphite samples is due to the contribution of a Hall resistance
generated by perpendicular current paths in the disordered media.
This appears to be the origin for the new phenomenon reported
here, namely the even QHE.

\ack
We thank M. Ziese for discussions. P.E. gratefully acknowledge
correspondence with M. M. Parish and Nai-Yi Cui.  The supports of the DFG
under Le 758/20-1, CNPq and FAPESP are gratefully acknowledge.




\begin{thebibliography}{00}
\bibitem{1} Y. Kopelevich, J.H.S. Torres, R. Ricardo da Silva, F. Mrowka, H. Kempa,
P. Esquinazi, Phys. Rev. Lett. {\bf 90}, 156402 (2003).

\bibitem{yakovadv03} Y. Kopelevich, P. Esquinazi, J. H. S. Torres, R. R. da
Silva, and H. Kempa, Adv. Solid State Physics {\bf 43}, 207
(2003).

\bibitem{ocana03}R. Oca\~na, P. Esquinazi, H. Kempa,
J. Torres and Y. Kopelevich, Phys. Rev. B{\bf 68}, 165408 (2003).

\bibitem{ulrich04}K. Ulrich and P. Esquinazi, J. Low Temp. Phys. {\bf 137}, 217 (2004).

\bibitem{luky04} I. A. Luk'yanchuk and Y. Kopelevich, Phys. Rev.
Lett. {\bf 93}, 166402 (2004).

\bibitem{matsui05} T. Matsui, H. Kambara, Y. Niimi, K. Tagami, M.
Tsukada, and H. Sukuyama, Phys. Rev. Lett. {\bf 94}, 226403
(2005).

\bibitem{2}J. Gonzalez, F.  Guinea, and M. A. H.  Vozmediano, Phys. Rev. B
{\bf 63}, 134421 (2001).

\bibitem{3}D. V. Khveshchenko,  Phys. Rev. Lett. {\bf 87}, 206401 (2001).


\bibitem{4}G. Baskaran and S. A. Jafari,  Phys. Rev. Lett. {\bf 89}, 016402 (2002).

\bibitem{041} K. S. Novoselov, A. K. Geim, S. V. Morozov, S. V. Dubonos, Y. Zhang, D.
Jiang, http://arxiv.org/find/cond-mat/0410631.

\bibitem{051}Y. Niimi, T. Matsui, H. Kambara, and H. Fukuyama, http://arxiv.org/find/cond-mat/0511733.

\bibitem{science}K. S. Novoselov, A. K. Geim, S. V. Morozov, S. V. Dubonos,
Y. Zhang, D. Jiang, Science {\bf 306}, 666 (2004).

\bibitem{novo05}K. S. Novoselov et al., Nature {\bf 438}, 197 (2005).

\bibitem{zhang05}Y. Zhang, Y.-W. Tan, H. Stormer and P. Kim, Nature {\bf
438}, 201 (2005).

\bibitem{carbon04}H. Kempa, P. Esquinazi and Y. Kopelevich, Extended
Abstract E012, Proceedings of the Carbon2004, International Conference on
carbon at Brown University, July 2004.

\bibitem{swm} J. C. Slonczewski and P. R. Weiss, Phys. Rev. {\bf 109}, 272
(1958); J. W. McClure, Phys. Rev. {\bf 108}, 612 (1957) and {\bf 112}, 715
(1958).

\bibitem{wal} R. R. Haering and P. R. Wallace, J. Phys. Chem. Solids {\bf 3}, 253 (1957).

\bibitem{5}A. A.  Abrikosov, Phys. Rev. B {\bf 60}, 4231 (1999).

\bibitem{6}M. M.  Parish and P. B. Littlewood, Nature {\bf 426}, 162 (2003).

\bibitem{7} H. L. St\"omer, J. P. Eisenstein, A. C.  Gossard, W.
Wiegmann, K.  Baldwin, Phys. Rev. Lett. {\bf 56}, 85 (1986).

\bibitem{8} S. T. Hannahs, J. S. Brooks, W.  Kang, L. Y. Chiang,
and P. M. Chaikin, Phys. Rev. Lett. {\bf 63}, 1988(1986).

\bibitem{9}S. Hill, S. Uji, M. Takashita, C. Terakura, T. Terashima,
H. Aoki , J. S. Brooks, Z. Fisk, and J. Sarrao ,  Phys. Rev. B
{\bf 58}, 10778 (1998).

\bibitem{10} D. Elefant, G. Reiss, C. Baier,  Eur. Phys. J. B {\bf 4},
45 (1998).

\bibitem{11}S. Uji, C. Terakura, M. Takashita, T. Terashima , H. Aoki, J. S.
Brooks, S. Tanaka, S. Maki, J. Yamada, and S. Nakatsuji,  Phys.
Rev. B {\bf 60}, 1650 (1999).

\bibitem{12}B.  Zhang, J.  Brooks , Z. Wang , J.  Simmons, J. Reno ,
N. Lumpkin, J. O'Brien, and R.  Clark, Phys. Rev. B {\bf 60}, 8743
(1999).



\bibitem{14} E. Abrahams, S. V. Kravchenko, M. P.  Sarachik,  Rev. Mod. Phys.
{\bf 73}, 251 (2001).

\bibitem{15}Y. Hanein, N. Nenadovic, D. Shahar, H. Shtrikman, J. Yoon,
C. C. Li, and D. C. Tsui,  Nature {\bf 400}, 735 (1999).

\bibitem{16} Y. Zheng and T. Ando,  Phys. Rev. B {\bf 65}, 245420 (2002).

\bibitem{gus05} V. P. Gusynin and S. G. Sharapov, Phys. Rev. Lett. {\bf
95}, 146801 (2005).

\bibitem{per05}N. Peres, F. Guinea and A. H. Castro Neto,
http://arxiv.org/find/cond-mat/0506709.

\bibitem{pauw}L. J. van der Pauw, Philips Res. Repts. {\bf 20}, 220
(1959).

\bibitem{17}H. W. Jiang, C. E. Johnson, K. L. Wang, and S. T. Hannahs, Phys. Rev.
Lett. {\bf 71}, 1439 (1993). M. Hilke, D. Shahar, S. H. Song, D.
C.  Tsui, Y. H.  Xie, and D.  Monroe,  Phys. Rev. B {\bf 56},
R15545  (1997).

\bibitem{18} H. Kempa. Ph.D. Thesis, Universit\"at Leipzig (2004),
unpublished.

\bibitem{19}A. A. Abrikosov,
Europhys. Lett. {\bf 49}, 789 (2000).

\bibitem{20} B. T. Kelly, {\it  Physics of Graphite}, Applied Science
Publishers LTD, London and New Jersey (1981).

\bibitem{21}A. C. Beer, {\it Galvanomagnetic Effects in Semiconductors},
Academic Press, N. Y. (1963).

\bibitem{22}L. Onsager,  Phil. Mag. {\bf 43}, 1006 (1952).

\bibitem{sta}F. A. Staas, A. K. Niessen, W. F. Druyvesteyn and J. v. Suchtelen, Phys. Lett. {\bf 13}, 293 (1964).

\bibitem{kop}Y. Kopelevich, V. V.  Lemanov, E. B. Sonin, and A. L. Kholkin, JETP Lett. {\bf 50}, 212 (1989).

\bibitem{vil} C. Villard, G. Koren, D. Cohen, E. Polturak, B. Thrane, and D. Chateignier , Phys. Rev. Lett. {\bf 77}, 3913 (1996).

\bibitem{bul} W. M. Bullis and W. E. Krag, Phys. Rev. {\bf 101}, 580 (1956).

\bibitem{bar} P. I. Baranskii et al., Phys. Status Solidi A - Applied Research {\bf 32},
K75 (1975).

\end{thebibliography}
\end{document}